\newcommand{\dd}[1]{\hbox{\rm #1}}
\newcommand{\VEV}[1]{\left\langle {#1}\right\rangle} 
\newcommand{\beq}[1]{\begin{equation}  \label{#1} }
\newcommand{\ceq}{\end{equation}}
\newcommand{\bear}[1]{\begin{eqnarray}  \label{#1} }
\newcommand{\cear}{\end{eqnarray}}

\newcommand{\hide}[1]{} 

\def\mnq{Mn$^{4+}$}
\def\mnt{Mn$^{3+}$}

\def\lcfifty{La$_{0.5}$Ca$_{0.5}$\-MnO$_3$}
\def\psfifty{Pr$_{0.5}$Sr$_{0.5}$\-MnO$_3$}
\def\nsfifty{Nd$_{0.5}$Sr$_{0.5}$\-MnO$_3$}

\def\ramo{R$_{1-x}$A$_x$\-MnO$_3$}
\def\lcmo{La$_{1-x}$Ca$_x$\-MnO$_3$}
\def\lc30{La$_{0.7}$Ca$_{0.3}$\-MnO$_3$}

\def\abs#1{\left| #1\right|}                
\documentstyle[epsfig,prl,aps]{revtex}
\begin{document}
\draft
\twocolumn 
[\hsize\textwidth\columnwidth\hsize\csname @twocolumnfalse\endcsname

\hide{ 
} 
\title{Field-induced segregation of ferromagnetic nano-domains in 
\psfifty, detected by $^{55}$Mn NMR} 

\author{G. Allodi, R. De Renzi, M. Solzi}
\address{Dipartimento di Fisica e 
Istituto Nazionale di Fisica della Materia,\\ 
Universit\`a di Parma, I-43100 Parma, Italy }
\author{K. Kamenev, G. Balakrishnan}
\address{ Department of Physics, University of Warwick, Coventry CV4 7AL, U.K.}
\author{M. W. Pieper}
\address{Institut f\"ur Angewandte Physik, D-20355 Universit\"at Hamburg, 
Germany. }
\date{\today}
\maketitle
\begin{abstract}
The antiferromagnetic manganite \psfifty\ was investigated at low temperature
 by means of magnetometry and $^{55}$Mn NMR. 
A field-induced transition to a ferromagnetic state is detected by 
magnetization measurements at a threshold field of a few tesla.
NMR shows that the ferromagnetic phase develops from zero field 
by the nucleation of microscopic
ferromagnetic domains, consisting of an inhomogeneous mixture of
tilted and fully aligned parts. At the threshold the NMR spectrum changes
discontinuously into that of a homogeneous, fully aligned, ferromagnetic
state, suggesting 
a percolative origin for the ferromagnetic transition.  
\end{abstract} 
]  

\narrowtext
Manganites \ramo\ (R = 
rare earth, A = alkali-earth metal) display correlated 
magnetic and transport properties, which include a 
colossal magnetoresistance (CMR) near $T_C$ for 
the metallic ferromagnetic compositions
(around $x = 1/3$ \cite{science}).
The complexity of the physics in manganites is witnessed by a very rich
phase diagram, which comprises various magnetic structures and
regions of phase coexistence at 
$x < 0.1$ \cite{hennion,manganese} and  at $x \approx 0.5$ 
\cite{Ca50,mori,calvani}.
The magneto-transport properties of these materials
are generally understood in terms of the double exchange interaction 
\cite{zener}, 
arising from spin-polarized carriers  coupled to 
localized electronic moments by a strong intra-atomic exchange. 
The underlying physics, however, is probably more complex,
and other competing interactions are relevant.
Among these, the narrow
bands, nesting effects of the peculiar Fermi surfaces, 
and the electron-lattice coupling through the Jahn-Teller (JT) 
active \mnt\ ion play perhaps a major role 
\cite{millis,mueller}.

Recently the focus of studies 
has moved to non-CMR compositions, 
in particular to the 50\% substituted compounds,
 where the itinerant ferromagnetic (F) state becomes unstable 
and  electronic localization with antiferromagnetic (AF) order take over 
at low temperature.
Manganites at half band filling 
display in fact two magnetically ordered states: 
a F metallic state at $T_c > T > T_N\approx150$K, 
and an AF insulating phase at lower temperature. The AF phase
 can be accompanied by the ordering of \mnt\ and 
\mnq\ on two distinct sublattices, like  in \lcfifty\ and \nsfifty 
\cite{marezio,kawano}.
We have recently shown \cite{Ca50} that 
in \lcfifty\ the charge ordered 
state sets in at $T_N$ 
by nucleation of mesoscopic AF domains from the 
ferromagnetic bulk in a first order 
transition.
However, in \psfifty\
charge ordering (CO) does not take place 
and the magnetic structure is of the layered A-type \cite{kawano}.
Both AF ground states 
can be destroyed by suitably strong applied fields, which restore the 
metallic F phase: 
this can be viewed as another regime of CMR.

In this paper 
we present an investigation of the 
AF-F transition in \psfifty , carried out by means
of $^{55}$Mn NMR, \hbox{a.c.} susceptibility, and magnetization measurements. 
The sample is a random 
assembly of small single crystals obtained 
from crushing a floating zone
specimen \cite{floatzone}.
Magnetization and a.c. susceptibility were measured by means of an
Oxford Instruments Maglab$^{2000}$ System 
($\mu_\circ H_{dc} = 0-7$ T, $T = 1.5-400$ K), equipped with  
a d.c. extraction  magnetometer and an a.c. induction susceptometer.
NMR was performed in liquid He at 1.3 K with a home built spectrometer 
\cite{pieper} and a variable field superconducting solenoid. 

\begin{figure}
\epsfig{figure=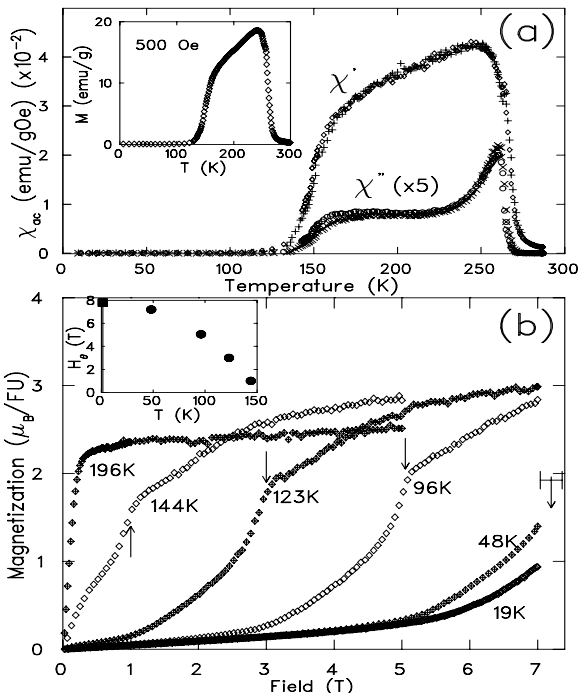,height=3.6in,width=3in,angle=0}
\caption{
a) Real, $\chi^\prime(T)$, and imaginary, $\chi^{\prime\prime}(T)$,
a.c. magnetic susceptibility. Measurements performed during cooling and 
warming are marked with closed and open symbols, respectively.
 Inset: $M(T)$ in 500 Oe. 
b) Magnetization $M(H)$ at several temperatures ($H$ swept from 0 to 7T). 
Inset: threshold field $H_\theta(T)$ (see 
text). $H_\theta(48\dd{K})$ is extrapolated from the data; the point
at 1.3 K 
is from the NMR measurements. 
}
\label{fig:pr50mag}
\end{figure}

A.c. susceptibility 
($H_{ac}=1$ Oe, $\nu = 1$ kHz)  in zero d.c. field was measured as a 
function of temperature.
The $\chi^\prime(T)$  and 
$\chi^{\prime\prime}(T)$ curves, 
reproducible on
cooling and warming, are shown in figure
\ref{fig:pr50mag}a.
The magnetization curve in a d.c. field of 500 Oe was 
measured as well, and it reproduces closely the features of $\chi'(T)$.
The curves show clearly the two magnetic transitions of the sample: the
Curie point $T_C= 270$ K, 
observed as a steep rise of $\chi'$ and 
the sharp peak on $\chi''$,
and the F-AF transition at $T_N \approx 150$ K, 
where the susceptibility drops by two
orders of magnitude.
This behavior is qualitatively similar to 
 that encountered in \lcfifty\   
where, however, in all reported works, a 
comparatively high remanent susceptibility (approximately 5-20\% of maximum, 
depending on the author) was found in the CO-AF phase. 
In the present case 
the susceptibility saturates below $T_N$ at the value  
$\chi' = 1.5\times10^{-4}$ emu/g\,Oe, 
only a factor 10 larger than expected in
a simple AF state by Curie-Weiss law, suggesting a very weak ferromagnetic 
term.
Moreover no appreciable thermal hysteresis was observed here,
in contrast again with \lcfifty\ 
\cite{Ca50,otherCa50}.  

Magnetization at constant temperature as a function of the applied field 
is shown in fig. \ref{fig:pr50mag}b for several 
temperatures below $T_C$. 
At $T_N < T < T_C$  an applied field of a few kOe fully saturates
the magnetization $M(H)$. 
Below $T_N$, the initial slope of $M(H)$ drops abruptly, 
corresponding to the onset of AF order.  
In both cases the initial d.c. susceptibility $\chi_{dc} = d M/dH$ is 
in good quantitative agreement with $\chi'_{ac}(T)$. 
In addition at $T < T_N$ a first order metamagnetic transition takes place at
larger fields: $M(H)$ deviates from 
the linear behavior, with a steep rise at a threshold 
field $H_\theta(T)$ (marked by arrows in the figure). 
This agrees with 
results from Tomioka et al. \cite{tomioka}.
The threshold field, determined as the knee of the curve, 
increases with decreasing temperature, as shown in the 
inset of fig. \ref{fig:pr50mag}b.
The magnetic moment 
right above $H_\theta(T)$ is roughly 2/3 of 
the apparent saturation value 
$\mu_s(T)\approx 3\mu_B$/formula unit,
which is approached at higher fields.
Such a large moment demonstrates that the magnetic order is close to that of 
a full ferromagnet above $H_\theta$. 


In order to obtain microscopic information on the field induced transition
we employed $^{55}$Mn NMR at 1.3~K in variable magnetic fields as a 
local probe of magnetization and of magnetic structure.
The local field experienced by $^{55}$Mn arises from dipolar,
transferred, and the Fermi contact fields, 
and is proportional to the electronic moment $g\mu_B\VEV{\hbox{\bf S}}$:  
\beq{eq:fieldmn}
\hbox{\bf B} = {{2\pi}\over{\gamma}}g\mu_B
{\cal A} \VEV{\hbox{\bf S}} + 
\mu_\circ\hbox{\bf H}. 
\ceq
Here 
$\mu_0{\bf H}$ is the external field, and $\gamma/2\pi = 10.501$ MHz/T for
$^{55}$Mn. The hyperfine coupling tensor $\cal{A}$ is found to be negative
and isotropic within the experimental resolution \cite{Ca50}. The resonance
frequency determines with this equation only the product ${\cal A}g\VEV{S}$. 
We use below the resonance frequencies in homogeneous Mn compounds as a
reference to assign local moments and a valence to different sites in our
spectra. 
The nuclei of the 3$\mu_B$ Mn$^{4+}$ ions 
resonate at low
temperatures around 300 MHz  
in several single valence insulating Mn compounds\cite{jaccarino,asai}. 
Similar frequencies have also been observed 
in CO manganites \cite{Ca50}. 
In the conducting CMR compositions $0.2\le x \le 0.5$, on the other hand,
the higher electronic spin $gS=4-x$ yields 
nuclear resonances at 1.3K ranging from 400 MHz down to 370 MHz 
\cite{Ca50,savosta,Mn4+}.

$^{55}$Mn NMR is also sensitive to 
local magnetic structure.
The superposition of the external and
the internal (hyperfine) field 
is different  in F and AF domains, giving rise to 
distinct shifts and broadenings for the 
corresponding resonance lines \cite{manganese,Ca50}. 
In particular in a F region, where Mn electronic
spins align parallel to the external field above 
saturation, the NMR resonance frequency shifts 
according to $\Delta\nu=g\mu_B\abs{{\cal A}} 
\VEV{S} - \gamma\mu_0 H/2\pi$, by eq. \ref {eq:fieldmn}.
Further information is provided by the radio frequency (rf) 
enhancement $\eta$, consisting of an amplification of both the effective 
driving rf field  $H_1^\prime=\eta H_1$ and the NMR signal induced in the 
coil, due to the hyperfine coupling of the electronic 
magnetization to the nucleus. The enhancement 
can be estimated from the rf 
power required for an optimized spin echo 
excitation \cite{eta}.
A large $\eta$ is typical of ferromagnetically ordered systems.

\begin{figure}
\epsfig{figure=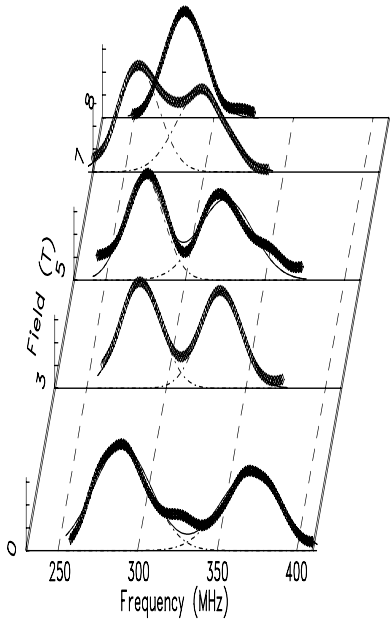,height=3.2in,width=2in,angle=90}
\caption{ $^{55}$Mn NMR spectra at 1.3 K for $0\le \mu_0H \le 8$ T. The 
spectra are rescaled in amplitude by arbitrary factors.  
}
\label{fig:pr50NMR}
\end{figure}

\begin{figure}
\epsfig{figure=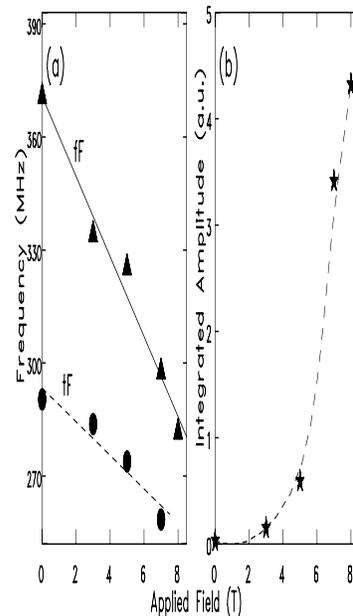,height=3.2in,width=1.8in,angle=90}
\caption{
a) Mean frequency of the fF and tF NMR peak as a function of the applied field.
Full line: slope $\gamma/2\pi=10.501$MHz/T; dashed: $\gamma/2\pi 
\cos(65^\circ)$ (see text).
b) Stars: integrated intensity $I(H)$ of the whole NMR spectra, corrected for 
the NMR sensitivity. The dashed line is a guide to the eye. 
}
\label{fig:NMRresults}
\end{figure}

The spin-echo spectra, measured at different
applied fields (always after zero field cooling), 
are plotted in fig. \ref{fig:pr50NMR},   
corrected for NMR sensitivity 
($\propto \omega^2$) and rescaled for clarity by 
arbitrary factors. 
The zero field spectrum (bottom of fig.\ref{fig:pr50NMR})
consists of a broad inhomogeneous distribution of hyperfine fields 
over a range of approximately 10 T,
where two peaks, centered  at 370 MHz and 290 MHz, are clearly resolved. 
In an applied field not exceeding 7 T, 
the whole spectrum shifts to lower frequencies, while the two peaks
get closer. This is evident from 
fig. \ref{fig:NMRresults}a, where the mean resonance frequencies,
from a two Gaussian components fit, are plotted 
as a function of $H$. 
Note that this spectrum differs from that 
of the F fraction of the CO manganite \lcfifty ,
consisting of a 
single much narrower peak (FWHM $\approx 2$ T) centered at 380 MHz.

Although the sample is antiferromagnetic at $H=0$ 
the high frequency NMR signal originates entirely from a ferromagnetic 
fraction,
as it is demonstrated by the sizeable enhancement 
$\eta\approx 100$ and by the field dependent frequency shifts. 
The slope of the full line in fig. \ref{fig:NMRresults}a shows
that the high frequency peak shifts with field according to the full nuclear 
gyromagnetic ratio ($\mu_\circ^{-1}d\Delta\nu/d H =\gamma/ 2\pi$, from
eq.\ref {eq:fieldmn}). This implies that the 
electronic moments on the Mn sites are constant and 
{\it fully} aligned to the external field,
as expected in a saturated ferromagnet.

The low frequency peak exhibits only a fractional shift 
(fig.\ref{fig:NMRresults}a), which implies 
a partial alignment of the Mn moments giving rise to this signal.
Assuming for the sake of simplicity a constant angle between external field 
and the Mn moments one
finds for this angle $\theta \approx 65$ degrees from the slope $d
\nu/\mu_\circ d H \approx \gamma \cos\theta/2\pi \approx 4.5$ MHz/T. 
We shall refer to this contribution as a {\it tilted} F (tF) component and
to the  former as fully F (fF).

Fig. \ref{fig:NMRresults}b shows the area  $I(H)$ under the full spectrum,
corrected for $\eta(H)$,
which is proportional to the number of resonating 
nuclei.
The  zero field signal  has a tiny intensity, hence the F fraction 
is initially a minority phase.
Its presence may account for the enhanced macroscopic d.c. susceptibility
in the AF state discussed above.
However, $I(H)$ increases rapidly with 
field, and the intensity ratio of the two peaks remains constant, of order one,
independent of the field.
The rapid increase of $I$ with field rules out an impurity phase.
No signal from the majority AF phase is observed,   
probably due to extremely fast relaxation,
as it is suggested 
by comparison with the related compound \lcfifty , where 
$^{55}$Mn relaxes two orders of magnitude faster in the AF phase than 
in the F phase \cite{Ca50}.

The field induced magnetic transition was easily located at 1.3~K and
7.7(1)~T by an abrupt change in $\chi$' and $\chi$''which induces a severe
detuning and rf-mismatch of the probe head.
The transition also shows up in the NMR spectrum, which  
at  8 T (the topmost in
fig. \ref{fig:pr50NMR}) consists  of a narrower single high 
frequency peak, whereas
the low frequency peak, still well resolved at  7 T, 
has completely disappeared. No additional peak was found between 210 
and 420 MHz.
The mean frequency of the 8 T spectrum lies on the same line of
the fF peak in fig. \ref{fig:NMRresults}a.
The tF peak is not recovered by setting the 
field back to 7 T: 
the hysteresis demonstrates that this is a first order transition, 
clearly corresponding to that detected 
by $M(H)$ at higher temperatures. 
Although instrumental limitations prevented 
direct verification by magnetometry, the value of 7.7 T at 1.3 K is  
in good agreement with the 
$H_\theta(T)$ curve extrapolated from magnetization data
(see inset to fig.\ref{fig:pr50mag}b).
The identification is also 
supported by the steep rise of the NMR amplitude $I(H)$
near 7 T, in qualitative  
agreement with the $M(H)$ curve at the lowest temperature (cfr.
fig. \ref{fig:NMRresults}b and fig. \ref{fig:pr50mag}b).

From our NMR data we can conclude, therefore, that on a microscopic scale
the increase of $M(H)$ below $H_\theta$ is not due to homogeneously
increasing induced moments or a field induced homogeneous canting of the
AF structure, since both are incompatible with the slope of the fF-line in
fig. \ref{fig:NMRresults}a. Instead, the simultaneous increase in the 
tF- and fF-line
intensities shows that $M(H)$ develops by inhomogeneous nucleation of fF-
and tF-phases from the AF matrix. 
The strong correlation between the intensity of the fF- and the tF-line
while both change with field by more than an order of magnitude strongly
suggests a growth of both phases in spatially connected volumes. It is
tempting to associate the two lines with the inner core and with the outer
surface layers of ferromagnetic clusters within the AF matrix respectively. 
At the threshold field the tilted
component vanishes, indicating that the ferromagnetic volume fraction
becomes homogeneous (AF-regions may still exist). 

If we follow this idea we may discuss some further consequences of our
data for the properties of these clusters. First, the fact that the
intensity ratio is constant means that the volume
fraction of the mixed phase (tF + fF) increases by growth in the number of
clusters rather than in their size. Second, the ratio
$I_{fF}/I_{tf}\approx 1$ implies a very large surface to volume ratio,
corresponding to a very small size of the clusters. Assuming for
simplicity a cubic shape, the core contains $(N-2)^3$ unit cells, covered
by a layer of $6(N-2)^2$ unit cells (not counting the edges). The NMR
intensity ratio then implies nearly equal volumes or $N=8$, that is a size
of the core in the order of six lattice constants. Finally, comparison
with the zero field frequencies of the reference materials described
above provides information on the local valence: 370~MHz for the fF-line
corresponds to 
Mn$^{+3.3 \le v \le +3.5}$ in a metallic ferromagnet, 
while 290~MHz for the tF-line is close to the value of Mn$^{+4}$ in 
antiferromagnetic insulators.

The peculiar nature of these clusters 
brings to mind a static version of  
magnetic polarons, often invoked by theories as the 
excitations of either magnetic JT \cite{dagotto} or magnetic
semiconductor \cite{nagaev} systems.
Unfortunately, we cannot distinguish from our NMR data between the two
cases of a tF core surrounded by a fF layer or vice versa, the ferromagnet
being surrounded by a tilted structure. From a magnetic point of view the
second possibility is more intuitive, but it implies some electrostatic
overshielding of the core hole state ($v \le +3.5$) in the surface
layer ($v\approx +4$), followed by the surrounding AF ($v\approx +3.5$). In
the other case the valence decreases 
nearly monotonically from the center of the
cluster where Mn$^{+4}$ forms an AF structure, canted due to the field and
frustrated magnetic bonds, to the fully aligned ferromagnetic surface of
the cluster. An interface layer between fF surface and the surrounding AF
matrix might well be unobservable in NMR. 
In both cases
the metamagnetic transition at $H_\theta$
indicates a change of topology in this phase. 
Its coincidence with a large mean magnetic moment
strongly suggests the crossing of a percolation threshold by F domains
at $H_\theta$. This view is also supported by the abrupt increase of 
electrical conductivity accompanying the transition \cite{tomioka}.

A similar intrinsic phase separation was encountered  in  \lcfifty , 
where a minority F fraction coexists with the majority AF phase at all 
temperatures below 150 K \cite{Ca50}. 
In that sample, however, 
the large thermal and magnetic hysteresis and the single 
fF peak in the $^{55}$Mn NMR spectrum indicate a bulk F phase. 
Recent TEM imaging actually showed that the size of F domains in \lcfifty\ is 
mesoscopic rather than nanoscopic \cite{mori}. 
In this respect \psfifty\ is more similar to low doped 
\lcmo\ ($x < 0.1$), where the nanoscopic dimension of 
spontaneously segregated  hole-rich F droplets  was demonstrated by small 
angle neutron scattering \cite{hennion}.
It is worth noting
that both \psfifty\ and under-doped \lcmo\ present the same A-type AF 
structure,  
\cite{hennion,kawano,wollan}, whereas the AF phase of  \lcfifty\ is charge 
ordered CE-type.  
Such a difference might be relevant: since the CO phase is far more 
insulating than the A-type phase \cite{kawano,tomioka}, the latter
may provide a screening mechanism sufficiently effective to cut the long
range tails of the Coulomb interactions and to accommodate     
charged clusters, whereas such a mechanism is ruled out in the insulating 
CO state.

In conclusion, 
$^{55}$Mn NMR in the AF state of \psfifty\ demonstrates the segregation 
of nanoscopic ferromagnetic clusters, dressed by a 
modulation of local spin and charge density at the 
interface with the host AF matrix. 
Evidence is provided that the field-induced transition to a 
ferromagnetic state, 
detected also by magnetization measurements,  is percolative in nature. 


This work was partially supported by  MURST-Cofin 1997 and 
 EPSRC GR/K95802 grants. Support by 
Prof. J. K\"otzler (IAP, Universit\"at Hamburg) is gratefully acknowledged.


\end{document}